\documentclass[twocolumn,a4paper,10pt]{extarticle}

\usepackage{mathptmx}

\usepackage[compact]{titlesec}

\titleformat{\section}[block]
  {\fontsize{14}{15}\bfseries}
  {\thesection}
  {1em}
  {}
\titleformat{\subsection}[block]
  {\fontsize{11}{15}\bfseries}
  {\thesubsection}
  {1em}
  {}

\usepackage[margin=0.75in]{geometry}

\usepackage{amsmath}
\usepackage{newfloat}
\usepackage{listings}
\usepackage{balance}

\usepackage[final]{microtype}
\usepackage[sort,nocompress]{cite}
\usepackage[usenames,dvipsnames,svgnames,table]{xcolor}

\usepackage{textpos}
 \usepackage{graphicx}

\usepackage{mathptmx} 
\usepackage{fancyhdr}
\usepackage[normalem]{ulem}
\usepackage[hyphens]{url}

\usepackage{enumitem}
\usepackage{dblfloatfix}

\usepackage{soul}
\soulregister\cite7
\soulregister\cref7
\soulregister\ref7
\soulregister\pageref7
\soulregister\itemize1
\soulregister\item1
\soulregister\ucache7
\soulregister\Ucache7

\usepackage{etoolbox}
\patchcmd{\section}{\scshape}{\bfseries\scshape}{}{}
\patchcmd{\subsection}{\itshape}{\bfseries\itshape}{}{}
\patchcmd{\subsubsection}{\itshape}{\bfseries\itshape}{}{}

\usepackage{newfloat}

\usepackage{booktabs}

\usepackage{flushend}
\PassOptionsToPackage{hyphens}{url}\usepackage[bookmarks=true,breaklinks=true,letterpaper=true,colorlinks,linkcolor=black,citecolor=blue,urlcolor=black]{hyperref}

\newcommand{\ucache}{filter cache}
\newcommand{\Ucache}{Filter cache}
\newcommand{\UCache}{Filter Cache}

\newcommand{\UCAche}{Filter Cache}
\newcommand{\ucacheh}{filter-cache}
\newcommand{\UCacheh}{Filter-Cache}

\usepackage{upgreek}

\newcommand{\hide}[1]{}

\newcommand{\myparagraph}[1]{\vskip .4ex\noindent\textbf{#1}\hspace{.6em}}

\usepackage{cleveref}

\crefname{figure}{figure}{figures}

\usepackage{subcaption}

\usepackage{tabulary}

\usepackage{changepage}   

\DeclareFloatingEnvironment[name={Attack},placement=t]{attackfig}
\crefname{attackfig}{attack}{attacks}
\newcommand{\attackcaption}{}
\newenvironment{attack}[1][Unspecified]{\renewcommand{\attackcaption}{#1}\begin{attackfig}\begin{adjustwidth}{.5em}{.5em}\rule{\linewidth}{.1pt}\vspace{0pt}}{\par\rule{\linewidth}{.1pt}\end{adjustwidth}\vspace{-3pt}\caption{\attackcaption{} Attack}\label{attack:\attackcaption}\vspace{-15pt}\end{attackfig}\par\noindent\ignorespacesafterend}

  \newcommand{\revision}[2]{#2}
  \newcommand{\urevision}[1]{#1}

\usepackage[square,sort,comma,numbers]{natbib}
\title{\huge MuonTrap: Preventing Cross-Domain Spectre-Like Attacks by Capturing Speculative State} 

\author{Sam Ainsworth \\  University of Cambridge \\  \textit{sam.ainsworth@cl.cam.ac.uk}
\and  Timothy M. Jones \\ University of Cambridge \\ 
\textit{timothy.jones@cl.cam.ac.uk}
}

\date{}

\begin{document}

\maketitle


\begin{abstract}

The disclosure of the Spectre speculative-execution attacks in January 2018 has left a severe vulnerability that systems are still struggling with how to patch.
The solutions that currently exist tend to have incomplete coverage, perform badly, or have highly undesirable edge cases that cause application domains to break.

MuonTrap allows processors to continue to speculate, avoiding significant reductions in performance, without impacting security.
We instead prevent the propagation of any state based on speculative execution, by placing the results of speculative cache accesses into a small, fast L0 filter cache, that is non-inclusive, non-exclusive with the rest of the cache hierarchy.
This isolates all parts of the system that can't be quickly cleared on any change in threat domain.

MuonTrap uses these speculative \ucache s, which are cleared on context and protection-domain switches, along with a series of extensions to the cache coherence protocol and prefetcher.
This renders systems immune to cross-domain information leakage via Spectre and a host of similar attacks based on speculative execution, with low performance impact and few changes to the CPU design.


\end{abstract}

\section{Introduction}

\noindent Speculative side-channel attacks, such as Spectre~\cite{Kocher2018spectre} and Meltdown~\cite{Lipp2018meltdown} have caused significant concern. While side channels, where secret data is leaked through unintended media to an attacker, have been well known and exploited previously~\cite{Chen:2010:SLW:1849417.1849974,SideChannelsPracticalLLC,Gruss:2016:PSA:2976749.2978356}, the wide applicability of the newer speculative attacks~\cite{Kocher2018spectre,Lipp2018meltdown,SpectrePrime} even to programs that are otherwise correct, the difficulty of fixing these side channels in software, and the high bitrates achievable through these speculative side channels, have resulted in a particular pressing need for good hardware solutions to remove the problem. 

The community is still grappling with how to deal with balancing the desire for performance, achieved through out-of-order execution and the relaxed microarchitectural guarantees it requires, against security properties only enforceable through in-order execution.
 Current software fixes~\cite{armfaq,kaiser,retpoline} either limit performance significantly, have limited coverage, or require security knowledge by the programmer.
 Existing solutions in hardware include restricting instructions that depend on speculative loads~\cite{8714070,NDA-micro,STT-micro}, which can work well for many compute-bound workloads but causes other workloads to suffer significant degradation.  Other techniques~\cite{yan2018invisispec} replay memory accesses at commit time, reducing throughput.

We argue that permitting the microarchitecture to continue speculating broadly on loads, which is necessary for the high performance of modern processors, must be a factor in any solution to speculative-execution attacks. 
We instead add in limited hardware regions where speculative hardware state can be observed, which can be cleared when there is potential for access by an attacker.
We design a speculative \ucache~\cite{Kin:1997:FCE:266800.266818}, which disallows propagation of speculative state into the rest of the system, including indirectly through the cache coherence protocol or prefetcher, preventing Spectre-like attacks between victim and attacker on a system.
This prevents leakage of information outside a protection domain, yet can be reused by many different actors on the same system with mutual distrust of each other.
Once a memory access becomes non-speculative its data can be placed safely in the main cache hierarchy.
However, if the loads that accessed the data are squashed, a cache line can remain securely in the filter cache until replaced via normal cache operation.

MuonTrap, our modest addition to a conventional out-of-order superscalar CPU, removes cache side-channels exploitable for speculative attacks, at low overheads (4\% slowdown for SPEC CPU2006, and 5\% speedup for Parsec). 



\section{Background}

\noindent Speculative side-channel attacks are possible because of a number of features of modern systems working together to create a vulnerability. We consider these here before describing currently implemented attacks in detail.

\subsection{Out-of-order Execution}

\noindent Almost all modern processors use some form of speculation when executing programs.
While this cannot affect the programmer's model, which should perform as though instructions are executed in-order, this does not prevent soft state such as in caches from being impacted by this execution.

While fetching and decoding of instructions is typically performed in order of the (speculative) program stream, their execution may be allowed to occur out-of-order, before then being retired in true program order. In an aggressively out-of-order core, if the data used by an instruction for a branch misses in the cache memory system, and the direction or target of that branch is mispredicted, then this misprediction may cause a large number of future, incorrect instructions to be executed before being thrown away.

This is important when hardware state can be impacted by the loading of secret data. If this secret data can be used as input to other instructions, we can indirectly leak it even if only accessed speculatively. This is particularly harmful on out-of-order processors, as they allow multiple instructions to reach execute before a mis-speculation is corrected.

\subsection{Timing Side Channels}

\noindent Side channels within a processor are a well-studied problem~\cite{Chen:2010:SLW:1849417.1849974,Chen:2014:ADC:2611765.2611766,Gruss:2016:PSA:2976749.2978356,Kocher:1996:TAI:646761.706156,SideChannelsPracticalLLC,Percival05cachemissing,Evtyushkin:2018:BNS:3173162.3173204}.
If execution on secret data can affect some indirectly visible property of the system, such as timing, then information about that data can be leaked without an attacker being able to directly access it. If we repeat this attack multiple times under different input scenarios, we may be able to leak the entire data item.

A particularly useful side channel is the memory system and its caches~\cite{Kocher2018spectre,Lipp2018meltdown,Liu:2014:RFC:2742155.2742176}.
Both the existence and non-existence of data in a cache can be used as a side channel: though the presence of cache data is not typically part of the programmer's model, by timing accesses we can observe information.
For example, by priming a direct-mapped cache with known data items, then allowing another process to load an address based on secret data, we can infer which element was evicted, and thus a subset of the loaded address.

Speculative-execution channels are particularly problematic because they allow us to introduce our own timing side channels into code, or access data that the programming model states isn't accessible, and therefore indirectly leak it. This means that, even if the programmer ensures any regions of code designed to access such data cannot leak timing information, an attacker can introduce new code to do so.






\subsection{Spectre}

\noindent Spectre~\cite{Kocher2018spectre} uses speculative execution on an out-of-order processor to leak secret data from a victim process to an attacker process, or between protection domains within a process.
It does this by tricking the victim into loading secret data speculatively, then using the result to do a second load with an address based on this secret.
This will evict data primed by the attacker from the cache, or bring in data shared between the victim and attacker, previously evicted by the attacker, whose access can subsequently be timed to leak bits of information. An example of this is shown in \cref{attack:Spectre Prime and Probe}.

 \begin{attack}[Spectre Prime and Probe]
	\includegraphics[width=\columnwidth]{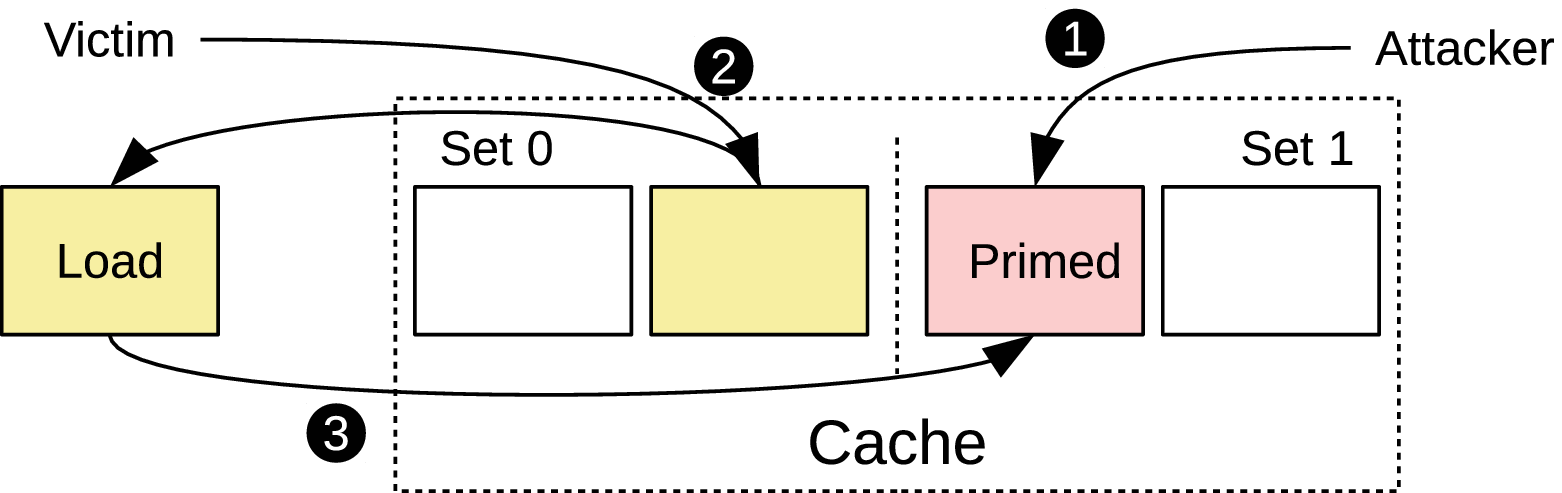}
	\vspace{-15pt}
	\myparagraph{Requirements} Shared cache between attacker and victim
	\myparagraph{Vector} Attacker brings in data into the cache \textbf{(1)}, followed by the victim being tricked into loading a secret value under speculation \textbf{(2)}, followed by a second load based on the contents of the first, which will or won't evict the primed data \textbf{(3)} depending on the secret's value.
	\vspace{-10pt}
 \end{attack}

There are two ways in which this attack can break down system barriers. One is between processes using inter-process communication: the attacker process can read the victim process's private data using such an attack. Within a process, in a low-level language such as C, user-space code can read any other user-space region, and so a Spectre attack to do the same is needless. However, if we have code in sandboxed regions, such as Javascript in a browser process, Spectre can allow the sandboxed code to implicitly read secret data outside of this region, unprotected by any kernel-level protection, and yet still considered harmful to leak into the sandbox.

\section{Requirements and Threat Model}

\noindent 
We seek to remove side channels introduced by the speculation, by moving speculative state into structures that can be cleared when necessary. 
Rather than preventing the use of any speculated or misspeculated state, much of which is both entirely innocuous and useful for performance, we instead consider it valuable to focus on a more precise threat model. We consider cache state affected by speculative execution to be vulnerable if either \textit{a)} a separate process is able to read information from it, or \textit{b)} in processes featuring untrusted code running in sandboxes (typically browsers), sandboxed code can observe speculatively accessed data from outside the sandboxed region but still within the process. 

This means that a user can potentially see metadata from their own speculative execution, but other attackers cannot.
An attacker is still, therefore, able to observe speculative execution by a victim if they can also trick the victim into timing a non-speculative access to the cache side channel the attacker has created before a context switch.
We assume that an attacker only has arbitrary control over a victim's speculative execution, and so such attacks cannot be executed.
\revision{B1}{Further, we are only interested in speculative side channels, rather than arbitrary covert channels---if the victim intentionally tries to send the attacker data, it will still be able to do so (for example, by deliberately affecting the timing of its own committed execution).
By making this simplification, we can allow the victim to observe timing characteristics of its own speculative execution, both before and after this misspeculation is rolled back by the processor.}

We only seek to remove speculative side channels from the memory system itself. In some microarchitectures other speculative side channels have been demonstrated, such as the clock when executing Intel AVX instructions~\cite{avxclock}, but these attacks do not involve hiding state to be picked up later, and do not involve potential chains of speculative behaviour, as is the case with cache loads, and so can be prevented by preventing soft state changes before speculation is completed.

We assume that protection within a sandbox, to prevent sandboxed code itself from speculatively reading other data, is achieved through other means, such as masked bounds checks on loads and stores~\cite{armfaq}.  This means we only have to focus on the more widely applicable and harder to prevent domain-crossing attacks such as between processes or through code within a sandbox calling code outside of the sandbox, avoiding slowdown for the vast majority of applications where sandboxed threat models do not apply, and avoiding unnecessary hardware overhead where possible.

Filter caches do not preclude the enforcement of stronger strategies that hide all information about the state of speculative execution that did not commit. But this simple policy is easy to enforce, does not require close coupling with the processor's internal state, covers the most interesting and widespread threats from Spectre-style attacks, and is permissive in terms of allowing optimisations where possible.

\section{MuonTrap}
\label{sec:ucache}

\begin{figure}
\includegraphics[width=\columnwidth]{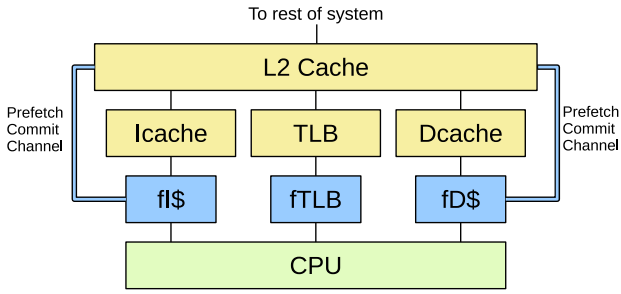}
\caption{The architecture of MuonTrap. \Ucache{}s are added for instructions, data and the TLB, which are cleared on a context switch, to isolate speculative data from being used for a speculative side-channel attack. Prefetch commit channels are added to preserve ordering of loads and stores without leaking speculative information, while still allowing larger caches to prefetch on data access.}
\label{fig:architecture}
\vspace{-10pt}
\end{figure}


\noindent We prevent speculative execution from leaving a trace that can be used for  side-channel attacks by adding a small, 1-cycle access L0 cache between the core and the L1 cache, which is cleared on context switches, kernel entry, and sandbox entry.
We force all speculative memory state to reside in the L0, meaning that other caches in the memory hierarchy contain only data that is non-speculative (i.e., it has been accessed by committed instructions only).
We call the L0 cache a \emph{speculative \ucache} and refer to other caches in the hierarchy as \emph{non-speculative caches}.

With MuonTrap, speculative memory accesses propagate through the conventional cache system, but do not evict data from non-speculative caches and do not alter data within them.
Data may be copied back to the L1 cache from L0 when a load or store using its cache line commits.
Although speculative data may be evicted from the \ucache{} before this point, it must not be written into a non-speculative cache. To prevent data from escaping, \ucache s are flushed on context switches, and between sandbox movement in processes with multiple actors in the same address space (such as JavaScript in the browser), and optionally on all misspeculation.
Adding this small cache increases lookup time in the L1 cache by one cycle, due to the need to consult the \ucache{} before the L1.
However, its size means it can be faster than a conventional L1 cache in an out-of-order superscalar system, so for memory accesses with high temporal or spatial locality, we should expect that this system may sometimes improve performance via hits to the \ucache, as well as provide security. 

In this section we first discuss the specification of and protection techniques employed in MuonTrap; the next section considers examples of specific Spectre-like side channels and how MuonTrap prevents their use.
The overall architecture of our scheme is given in \cref{fig:architecture}.

\subsection{\UCAche}

\noindent We use the \ucache{} to isolate speculative data from the existing cache hierarchy, while still providing access to this data to other speculative instructions to improve performance. This means that higher levels of cache are not filled upon a miss (speculative or otherwise); instead this data is brought directly into the \ucache. 
Ideally, the \ucache\ should be large enough to store the volume of speculative data that exists in common execution traces (i.e., for the maximum likely number of speculative loads and stores in the load and store queues, \revision{B3}{where we assume stores can prefetch cache lines from memory into the \ucache{}, but cannot perform an exclusive read until commit}).
Otherwise, data will be evicted from the \ucache{} before it is committed, and won't reach the L1 cache, limiting the temporal locality we can exploit, as the data must be reloaded on next use. 

\begin{attack}[Inclusion-Policy]
\includegraphics[width=.95\linewidth]{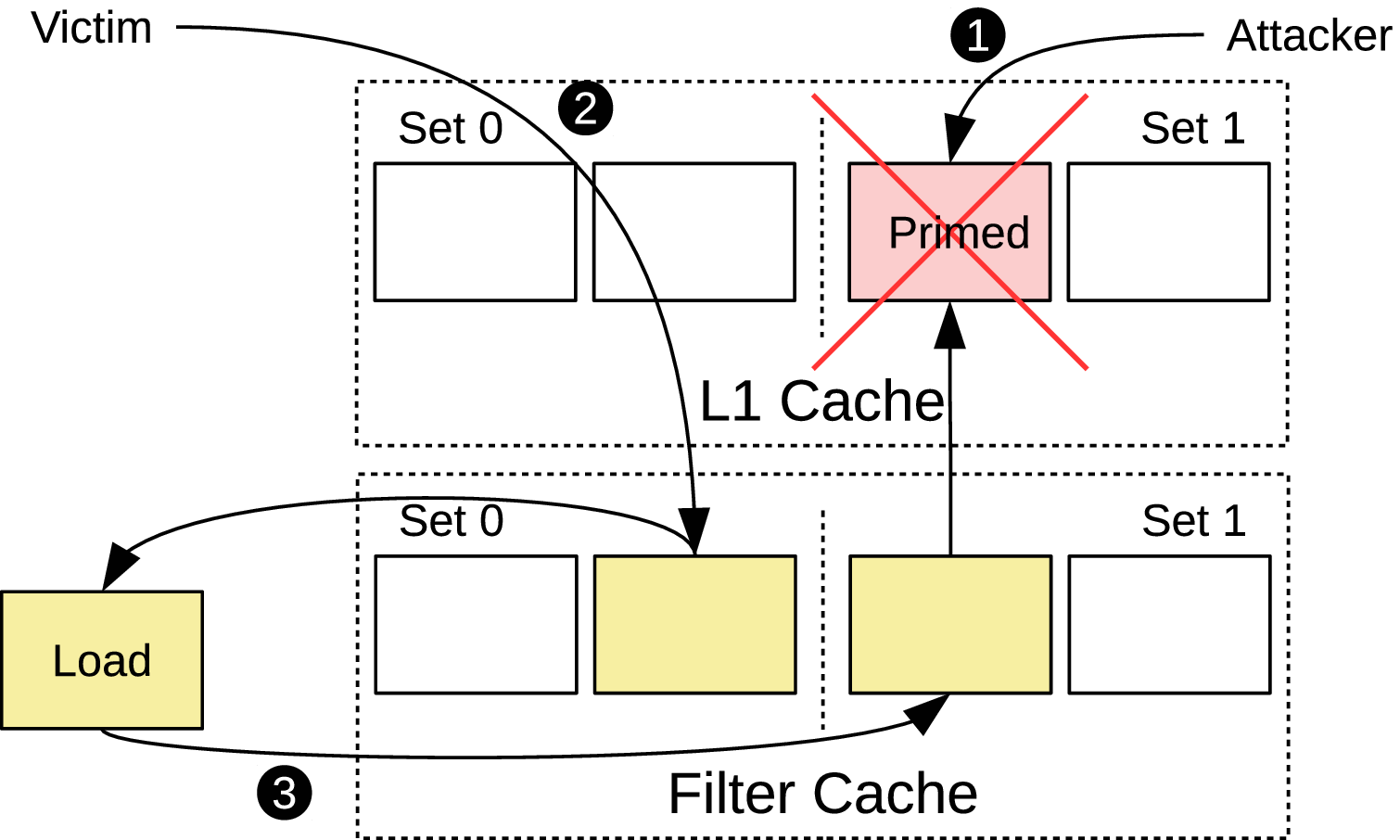}
\myparagraph{Requirements} Inclusion or exclusion with shared data between attacker and victim
\myparagraph{Vector} Priming the L1 cache \textbf{(1)}, followed by loading of secret data \textbf{(2)}, and using that to evict data indirectly from the L1 by using inclusion \textbf{(3)} or exclusion with shared data
\myparagraph{Defense} Non-exclusive, non-inclusive \ucache
	\vspace{-8pt}
\end{attack}

A speculative \ucache\ is non-inclusive non-exclusive with respect to the rest of the system's caches. To see why, consider \cref{attack:Inclusion-Policy}: we must prevent data brought into a \ucache\ from influencing state anywhere else in the system, and so inclusion and exclusion must both be prohibited. This means that, as with an exclusive cache, data propagates up the non-speculative cache hierarchy in reverse. It is brought into the L0 \ucache, and when committed it is written back out to the non-speculative L1 (and higher caches in an inclusive hierarchy), becoming visible to the rest of the system. 
We assume that cache-line sizes are the same at all levels of the cache hierarchy, so data from the \ucache\ can be used to fill any other cache.
If cache-line sizes differ, the \ucache\ must take the size of the largest cache line in the system, and write through to this cache upon eviction of data.

The \ucache\ itself is faster than a moderately sized L1, provided the data is in the \ucache.
This is because it is a small cache that can be virtually tagged and addressed from the CPU-side, and only ever contains data from one process, as it is flushed on context switches.

\subsection{Cache-Line Commit}

\noindent We add a \emph{committed} bit to each \ucacheh{} line, which is set to zero when a cache line is brought in by a speculative instruction, or set to one when a cache line is brought in through a non-speculative instruction. This means that an uncommitted line (i.e., containing speculative data) will not be written back to the L1. 
When memory accesses reach in-order commit in the out-of-order pipeline, the cache is accessed and the \emph{committed} bit for the cache line is set if it is zero, and the line written through to the L1 cache, regardless of whether the operation was a read or write.
It is left in the L0 to improve hit time. 
This write-through-at-commit policy increases the performance of cache flushes, since all data in the \ucache\ can safely be thrown away at any point during execution.  
Data only propagates into non-speculative caches if an instruction using that data has reached commit in the out-of-order processor pipeline; that is, if the data would have been loaded in the absence of speculative execution. 

When multiple speculative instructions use the same cache line, if any commits then the relevant cache line should be written through to the L1 cache and the L0 line marked as committed. If the line is no longer in the L0, then it is requested again from the rest of the memory system, and brought into the L1. 
This is because a valid in-order execution would also have brought this data into the cache, and so the state should become observable to the rest of the system. Even if subsequent instructions using the same line do not commit,  the cache lines that should be in the L1 do not change, and so an attacker cannot observe any further information from this.

\subsection{\UCacheh\ Clearing}

\noindent A \ucache\ is cleared upon a context switch, system call or protection-domain switch, to prevent leakage of data via its presence or absence within the speculative \ucache\ between protection domains. As we need not write back any data upon a flush in this write-though \ucache, we can simply invalidate all data to make it invisible in the hierarchy.  

We implement cache invalidation efficiently by storing a \emph{valid} bit per cache line, in registers separate from the SRAM cache. 
On a context switch, rather than having to access every SRAM block, which may take as many cycles as there are cache lines, we can invalidate the entire cache by clearing every valid bit, which can be performed in parallel within a single cycle.
On lookup, any cache line with the valid bit unset is ignored in the \ucache.
This is unconventional for caches, which normally store validity implicitly using coherency state in SRAM, but is necessary for fast invalidation of the cache, and the extra state is feasible considering the small size of a \ucache.
It is this fast invalidate that requires the \ucache{} to be write-through.

Note that we do not flush the \ucache\ on mispredicted branches.
This is because many applications make use of data loaded in on mispredictions to improve performance, as such branches are likely to be taken in the future. Since this does not cause cache-timing leakages to other protection domains, we leave this data in the \ucache\ except from on context switches, system calls or other protection-domain switches, when all committed and uncommitted data is cleared.

 \begin{figure}[t]
 \begin{center}
  \includegraphics[width=.8\columnwidth]{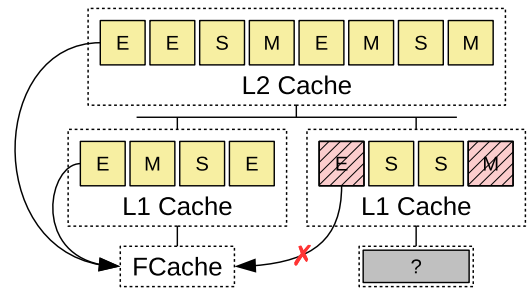}
 \end{center}
  \vspace{-10pt}
 \caption{\revision{C1}{A \ucache\ can have shared access to any data in caches on the linear path to memory, and any data in shared state in private caches, but not data in modified or exclusive in other parts of the hierarchy. This lets the \ucache\ speculatively read the data without affecting any non-speculative coherence states or viewing the contents of other \ucache s.}}
 \label{fig:ucachecoher}
 \vspace{-10pt}
 \end{figure}
 
\subsection{Addressing}
\label{sec:ucache:addr}
\noindent The \ucache\ is accessible in a single cycle, so it is desirable to avoid virtual-address translation on access.
Clearing it on a context switch avoids aliasing between different physical addresses for shared virtual addresses across multiple processes.
However, the \ucache{} must be checked by the cache-coherence logic, so it must be possible to index it by physical address.
We therefore tag each entry with both the virtual and physical address of the data and index the cache by the shared least significant bits of both.
This means it is virtually indexed from the CPU side and physically indexed from the memory side, to avoid translation.
We also prevent virtual-address aliasing within a process by physically addressing upon a memory fill, which may overwrite an alias with a different virtual address but ensures that only one copy of each physical address exists at a time in the \ucache.

\subsection{Coherency Mechanism}
\label{ssec:coherency}
 \label{sssec:coherenceprevention}
 
\noindent \revision{C1}{\Ucache s can only participate in the cache coherence protocol in a way that cannot be timed by non-speculative caches, does not affect the state of any non-speculative caches, and does not influence the behaviour of other \ucache s in the system.
To achieve this, using the MESI protocol, we allow any \ucache\ to hold a copy of some data in shared (S), provided that this data is only in an exclusive (E) or modified (M) state in a non-speculative cache nearer, and on a direct path to, main memory, or no non-speculative cache has a copy in E or M state.
This is exemplified in \cref{fig:ucachecoher}.
Upgrades from S, when a load or store is written through to the L1, can only occur once the associated instruction is non-speculative, and must invalidate any other \ucache\ that may have a copy under this policy in a constant-time operation.}

\urevision{What follows is that, if data is held in a non-shared way in a cache that is private to another part of the hierarchy, then speculative accesses must wait to access the data until they become non-speculative.
Likewise, for all \ucache s to work independently and not leak speculative information to each other, \ucache s cannot enter non-shared states.}



 \begin{attack}[Shared-Data]
 \includegraphics[width=\linewidth]{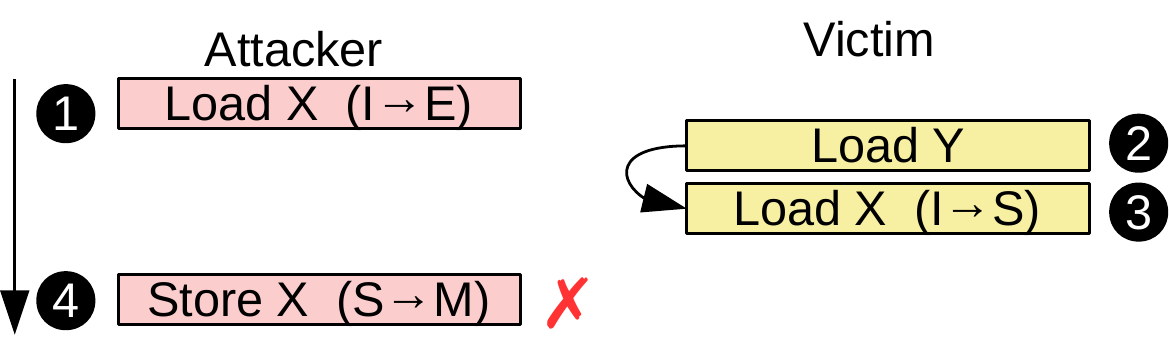}
 
 \myparagraph{Requirements} Shared data between the attacker and victim, with write access for the attacker, or ability for the victim to perform speculative prefetching in the exclusive state or issue speculative-write coherency requests~\cite{SpectrePrime}
 
 \myparagraph{Vector} Priming the cache \textbf{(1)} with a load, followed by tricking the victim into loading a secret \textbf{(2)} and using that to trigger a load \textbf{(3)} or store attempt~\cite{SpectrePrime}, increasing the time taken for the attacker to perform a store \textbf{(4)}
 
 \myparagraph{Defense} Reduced coherency speculation
 	\vspace{-8pt}
 \end{attack}
 
\myparagraph{Reduced Coherency Speculation}
We prevent \ucache s from changing the coherence state of any non-speculative cache via speculative instructions.
If a private cache on a different part of the hierarchy is in M or E state, then bringing this data into S state within the \ucache\ would change the other cache's state.
\revision{C1}{By comparison, an access by the \ucache\ to main memory, or to a cache closer to memory than the L0, would not, as any filter cache below this is allowed a copy of the data.}
\Cref{attack:Shared-Data} shows an example mechanism.

To prevent such an attack, during speculative execution we delay any memory access that would cause another private cache to move from M or E into S or I, until it is at the front of the instruction queue, and so will definitely be executed within the program.
\revision{B4}{Any operation that is affected by this delay is negatively acknowledged to the requesting core, which repeats the access when safe.
Forward progress is maintained, as any coherence transaction that reaches the head of the queue will succeed.}
As these delays are purely based on the contents of non-speculative caches, they cannot leak speculative data.

\myparagraph{\UCacheh{} State Reduction}
We further ensure that any speculative access from one \ucache\ has independent timing characteristics from any other, even after commit. This is by allowing data to only be brought into the \ucache\ in Shared state. This means that the presence of data in any other \ucache\ is not leaked by the coherence protocol, even after any relevant instructions commit and are sent into the non-speculative cache hierarchy, because with this constraint we only need to check non-speculative caches to move into the shared state. Indeed, as the \ucache\ is write-through, there is no direct need for any states other than shared (S) and invalid (I) within a \ucache.  
Without this reduction of possible states, and associated timing guarantees, then attacks such as \cref{attack:\UCacheh{} Coherency} would be possible, both before a speculative load is squashed, and given that \ucache s are not flushed on a misspeculation, after correct execution is restarted. 

 \begin{attack}[\UCacheh{} Coherency]
 \includegraphics[width=\linewidth]{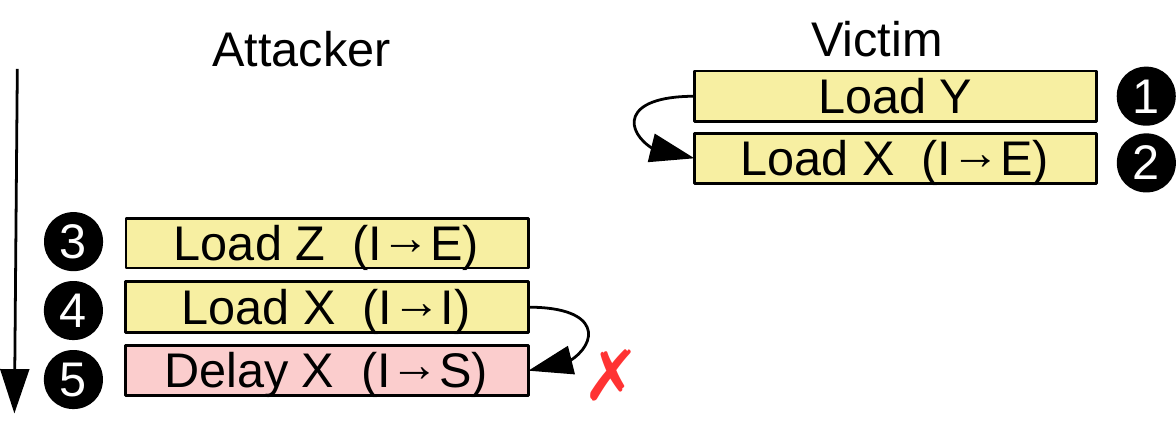}
 \myparagraph{Requirements} \Ucache\ with reduced coherency speculation, and shared data between the attacker and victim
 
 \myparagraph{Vector} Speculative load access by the victim to secret data \textbf{(1)} used to trigger an access to shared data \textbf{(2)}. This is followed by a load to an address the attacker knows will miss \textbf{(3)} causing a load to the same address accessed speculatively  by the victim \textbf{(4)} to be delayed \textbf{(5)}
 
 \myparagraph{Defense} Loading into the \ucache\ as shared, with asynchronous upgrade at commit (for performance)
 	\vspace{-8pt}
 \end{attack}

This could effectively reduce behaviour to MSI, even if MESI is internally supported, because if we assume all loads are speculative, no data will ever be brought in to E state.
That said, states such as O and F in MOESI and MESIF will be used, as instructions entering those states would become non-speculative.
Yet MESI, or more complex protocols, are typically supported in modern processors for performance reasons, and so we want to have data in exclusive state whenever useful.
To achieve this efficiently, we add a new pseudo-state to the \ucache, SE, which behaves like S to the coherence protocol, but when the load becomes visible to the rest of the system, an asynchronous upgrade to the E state is launched from the L1 cache, similar to a prefetch. \revision{C3}{From a functional protocol perspective, this means we add no new states; an L0 can only take on the S or the I state.}
A line in the \ucache\ is placed in SE when an unprotected system would have placed it in E in the L1; that is, the data is in no other private non-speculative cache in the system. The asynchronous upgrade this triggers at commit invalidates any copies in the rest of the hierarchy
written back since, and in any \ucache{}s, which remain invisible from timing.

\myparagraph{Wider Implications}
These constraints allow \ucache{}s to participate in the cache-coherence protocol without any visible timing effects from speculative execution.
This means that we need not repeat memory accesses, and can store data brought in by a \ucache\ immediately into the non-speculative cache hierarchy.
As an effect, a commit of a load cannot be stalled by any second cache access, unlike in previous work~\cite{yan2018invisispec}.
\revision{A2/B7}{Further, it means that memory accesses that do not require state changes in caches private to other cores can continue to be arbitrarily speculated and executed in parallel, unlike techniques that restrict L1 misses in general~\cite{SDVP}.}
\revision{B3}{While a speculative store instruction cannot prefetch data into E state in the \ucache, it can still bring data in from closer to memory in S state, speeding up the write post-commit.}

The cost of allowing \ucache s to participate in coherency is that upgrades to exclusive or modified must invalidate other \ucache s in the rare event that the data isn't already in an exclusive state within a cache private to the upgrading core. 
\revision{B5}{This broadcast is designed to provide timing invariance with respect to data in filter caches, which must be preserved even if a snoop filter is used. 
Still, this is unnecessary in the typical case, where we already have exclusive access in the L1 cache. More generally, the technique scales to many-core hierarchies by only requiring multicast to cores in clusters below a shared cache with exclusive access to the data, as these are the only \ucache s which could currently store shared copies of the cache line.}



MuonTrap does not directly interact with the memory consistency model. All it is allowed to do in this respect is delay load operations, but this neither makes any particular ordering stricter nor weaker, and so any commonly-used consistency model can directly use MuonTrap.

\subsection{Prefetching}
 \label{ssec:prefetcher}

  \begin{attack}[Prefetcher]
  \includegraphics[width=\linewidth]{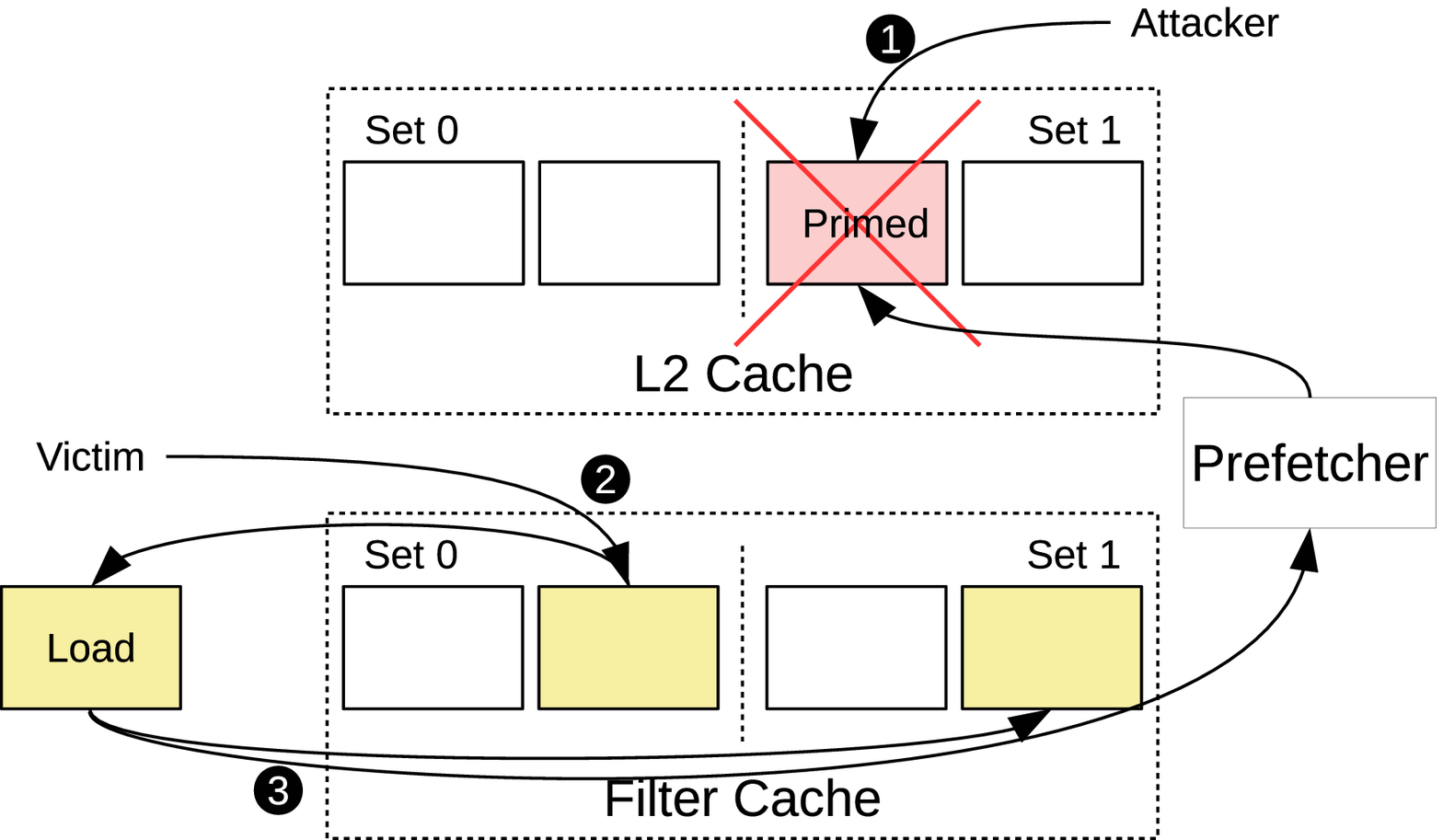}
  	\vspace{-18pt}
  \myparagraph{Requirements} Prefetcher with access to speculative loads
  \myparagraph{Vector} Priming a cache featuring a prefetcher \textbf{(1)} by the attacker, followed by loading secret data \textbf{(2)} by the victim, who then loads an address to trigger the prefetcher to bring the next line into a non-speculative cache \textbf{(3)}
  \myparagraph{Defense} Prefetch on commit
  	\vspace{-8pt}
  \end{attack}

\noindent It is insufficient to hide speculative memory accesses in \ucache{}s if those accesses can indirectly trigger changes in non-speculative cache state from prefetches based on them. To see why, consider \cref{attack:Prefetcher}.
By bringing in further lines based on the speculative access, a prefetcher would leak information to the wider hierarchy. We must therefore trigger prefetches only based on the committed instruction stream, rather than speculative accesses. 
We  add a tag to each cache line in the \ucache, specifying which level of the non-speculative hierarchy it was brought in from. When a \ucacheh\ line changes state from uncommitted to committed, a prefetch notification is then sent to the corresponding level, provided it has a prefetcher, to avoid triggering unnecessary prefetches to caches that weren't accessed. 


\subsection{Protected Caches}

\noindent Though this section has so far assumed that the \ucache\ is used to protect data-cache loads, this is not the only place in which a \ucache\ is needed in a system to prevent speculative side-channel attacks.


 \begin{attack}[Instruction-Cache]
 \includegraphics[width=.95\linewidth]{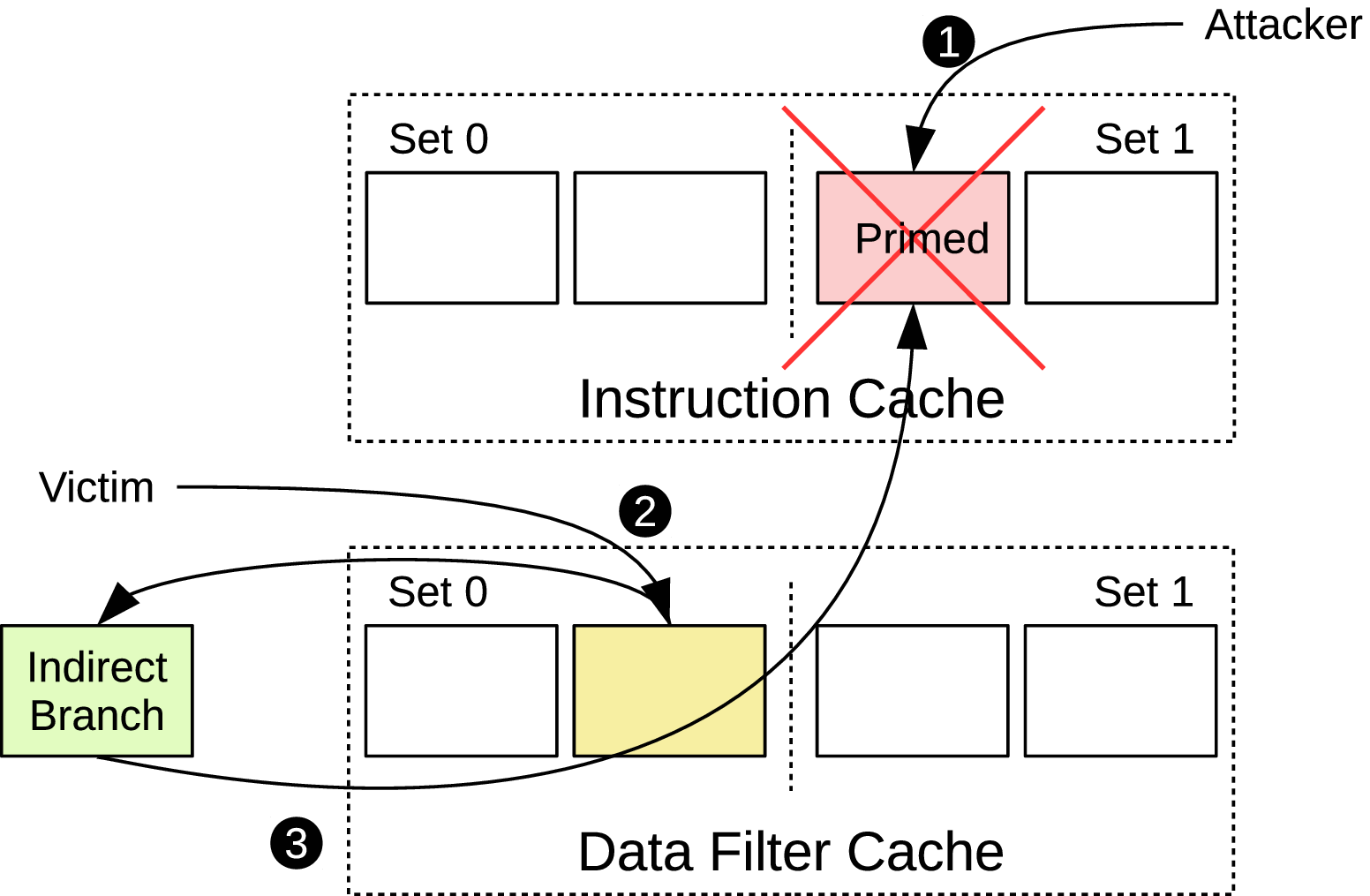}
 	\vspace{-5pt}
 \myparagraph{Requirements} Inst. cache shared by attacker and victim
 \myparagraph{Vector} Priming the instruction cache \textbf{(1)} by the attacker, followed by loading of secret data \textbf{(2)} and using it to index into the instruction cache using an indirect branch \textbf{(3)}
 \myparagraph{Defense} \Ucache\ for instruction access
 	\vspace{-8pt}
 \end{attack}

\myparagraph{Instruction Cache}
\noindent Though the instruction cache was not used as an attack vector for the original Spectre attack~\cite{Kocher2018spectre}, it is possible to leak speculative information in a similar manner to attacking the data cache. For example, in \cref{attack:Instruction-Cache} 
an attacker can cause a victim process to jump to a memory location based on the value of secret data, and then infer this information by timing instruction access. 
  We can fix this side channel in the same way that we fix the data side channel: by using a speculative \ucache\ for instructions in addition to that for data.
  This is comparatively simpler because there is no cache coherency for read-only data, and so we only need to set the committed bit on committing an instruction, without issuing any further memory transactions to upgrade the cache line.

\myparagraph{TLBs and Page-Table Walkers}
\label{ssec:utlb}
\noindent Side channels within a TLB are more difficult to exploit cross-process than on other caches, as no shared TLB translations will exist.
Therefore attackers are restricted to prime and probe attacks, where they infer data by causing the victim process to evict translations placed in the TLB by the attacker.
Still, for full protection we need to prevent eviction of non-speculative entries with speculative translations, by storing speculative translations in a filter TLB. On instruction commit, the relevant translations are moved to a nonspeculative TLB, and the filter TLB is flushed on a context switch, as with a \ucache. 


Hardware page-table walks triggered from speculative instructions can write into caches.
We must also  prevent side channels through this mechanism.
Although we can enforce that these memory accesses go through a \ucache, working out when a page-table-walker cache entry should be committed is complex, as there is no one-to-one correspondence between executed instructions and page-table entries.
Upon commit, we retranslate any instructions that caused page-table misses: these entries are likely to be in the \ucache\ at this point, and can be written back to the L1 as a result of the non-speculative retranslation. This allows us to mark \ucacheh\  elements that should be written through into the L1, because they were part of valid non-speculative execution.

\subsection{Multicore and SMT}

\noindent Different cores may have different processes running within them at any one time, and so each must have a \ucache\ to isolate them from each other's speculative side channels. Assuming that they are allowed to be from separate processes, this is also true of multiple threads in a simultaneous multithreading arrangement on the same core. Protection of the \ucache{} relies on isolating speculative data within the \ucache{} from threads from other processes.
This necessitates that when multiple processes are run concurrently via simultaneous multithreading on the same core, they must not be able to infer \ucacheh\ state about the other, otherwise information may leak between the two.
This means each thread must similarly have either separate \ucache s, or use partition-based isolation based on the process ID.

\subsection{Within-Process Attacks}

\noindent MuonTrap prevents between-process attacks by clearing the \ucache\  on a context switch, and user-kernel attacks by clearing on syscalls, thus preventing the attacker from learning any information based on what has or hasn't been loaded.
However, an attacker may be in the same process as the victim, executing within a sandbox (if not inside a sandbox, the attacker already has nonspeculative access to the victim's data).
We therefore need to clear \ucache s on movement into sandboxed regions. This is performed using a dedicated flush instruction sat behind a non-speculation barrier~\cite{armnonspec}.
As a result, speculation barriers only need to be inserted at the boundaries of sandboxes, rather than throughout the entire program as is currently necessary. This defense requires that an attacker can only cause execution outside of the sandbox by a path terminated with a \ucacheh\  clear. For full protection against the subset of variant 2 attacks where the victim is fooled into mistraining its own branch targets by an attacker sandbox, branch-target-buffer isolation, as is already implemented on recent commodity systems\cite{armv85aupdates}, is also necessary. For protection against Spectre 1.1~\cite{kiriansky2018speculative} within a sandbox, where speculative stores are used to overwrite return stacks and thus execute code outside the sandbox, stores within sandbox-interpreted code should be covered by  masked bounds checks~\cite{armfaq}, and to prevent the sandboxed code itself from executing Spectre v1 attacks, masked bounds checks should also be used on sandbox-interpreted loads. By utilising simple software fixes where applicable, we can avoid the overheads of hardware mitigation for the vast majority of applications that are not vulnerable to such attacks, while covering those which are hardest to fix by using dedicated hardware. 

\revision{A1}{Still, for applications requiring strong protection properties of their sandboxes that are not provided in software, we provide the option to \textit{clear-on-misspeculate}, on a per-process granularity. This prevents the timing of a speculative side channel after execution restarts.}

\subsection{Remaining Channels}

\revision{E1}{MuonTrap follows a permissive threat model---its main goal is to prevent speculative side channels between distinct domains, such as process boundaries, rather than arbitrary speculative side channels.
This manifests in multiple ways:}

\myparagraph{Data Visibility}
\urevision{By default, MuonTrap does not clear its filter caches upon misspeculation, though this can be enabled per-process.
Without this clearing, speculative state in the filter cache can be observed by the victim themselves after correct execution is resumed.
However, this is not passed on to attacker-run code, and it allows any data brought in under normal misspeculation to be reused.}

\myparagraph{Contention}
\urevision{MuonTrap does not allow speculative \ucache{} state to affect any non-speculative cache.
Their interaction only occurs at commit time, with the results of loads and stores being written from the \ucache\ to L1 in program order, provided new data was brought in.
Still, an under-sized filter cache can have speculative data evicted from it before commit-time.
In this case, the data will be passively re-loaded from memory or a higher cache level into the L1, causing transient contention in caches.
Since this requires arbitrary control of both the victim's committed and speculative execution, and only causes a transient channel, we consider it out-of-scope.}

\myparagraph{Timing}
\urevision{Even with clear-on-misspeculate, MuonTrap does not prevent code that will be committed from observing the timing behaviour of concurrently running speculative execution within the same thread~\cite{fustos2020spectrerewind}.
It shares this property with any techniques that allow wider propagation of speculation and roll it back~\cite{Saileshwar:2019:CUA:3352460.3358314}, but not with those that limit speculation~\cite{8714070,STT-micro,NDA-micro}.
Still, MuonTrap mitigates by preventing concurrent execution from multiple threat domains within the same thread by using barriers and flushes, and by having a constant-time \ucache\ flush operation, rather than an undo that depends on the speculative actions it has performed.}

\subsection{Summary}

\noindent To prevent speculative data from being leaked within the memory system, we have added MuonTrap, an extra layer of indirection, into the cache hierarchy.
This \ucache{} sits between each core and the L1 caches to store speculative data and is cleared on switches of protection domain.

A basic \ucache\ in front of the L1 data cache, while able to defend against traditional Spectre attacks, can still allow information leakage through the cache hierarchy. We must therefore impose further constraints on the cache coherence protocol, add \ucache s to the TLB system and instruction cache, and prevent data leakage via the prefetcher, for our system to be secure against other similar attacks.

\section{Experimental Setup}

\noindent 
We model a high performance multicore system using the gem5 simulator~\cite{Binkert:2011:GS:2024716.2024718} with the ARMv8 64-bit instruction set and configuration given in \cref{tab:setuptable}, similar microarchitecturally to systems used in previous Spectre mitigation techniques~\cite{yan2018invisispec,STT-micro}. The L1 data cache is 2-cycle access, and the L1 instruction cache 1-cycle: this is because the instruction cache is typically faster in a modern system~\cite{SourcesOfError}, as it can be virtually tagged without aliasing issues. We simulate SPEC CPU2006~\cite{SPEC2006} in syscall emulation mode, fast forwarding for 1 billion instructions, then running for 1 billion instructions. We also evaluate on Parsec~\cite{Parsec}, running on the simsmall datasets with 4 threads in full-system mode running Linux. The benchmarks included are the subset that compile on AArch64, and are run to completion. Results for the open-source InvisiSpec~\cite{yan2018invisispec} and STT~\cite{STT-micro} are reproduced on the same AArch64 system, using recent patches~\cite{invisifix,stt-repo}. 

\begin{figure*}[t]
	\centering
	\includegraphics[width=2\columnwidth]{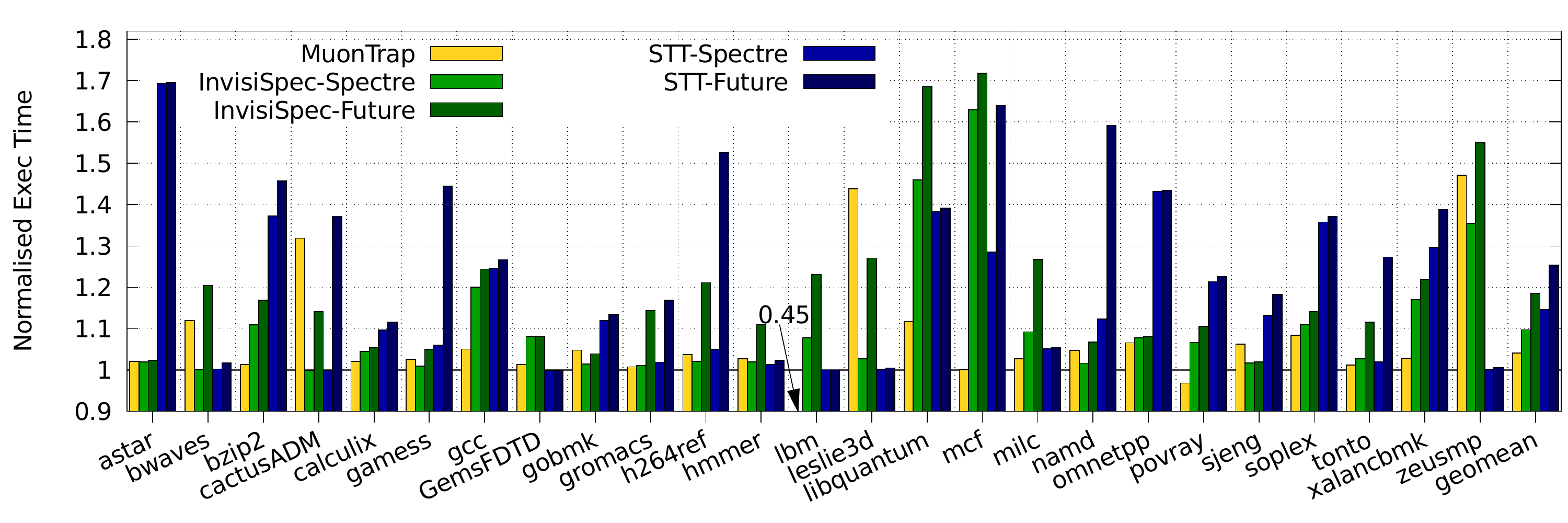}
		\vspace{-10pt}
	\caption{Normalised execution time from MuonTrap, compared with each of the InvisiSpec~\cite{yan2018invisispec} and STT~\cite{STT-micro} techniques running SPEC CPU2006 (lower is better).}
	\label{figs:spec2006}
	\vspace{-15pt}
\end{figure*}

\begin{figure}[t]
	\centering
	\includegraphics[width=\columnwidth]{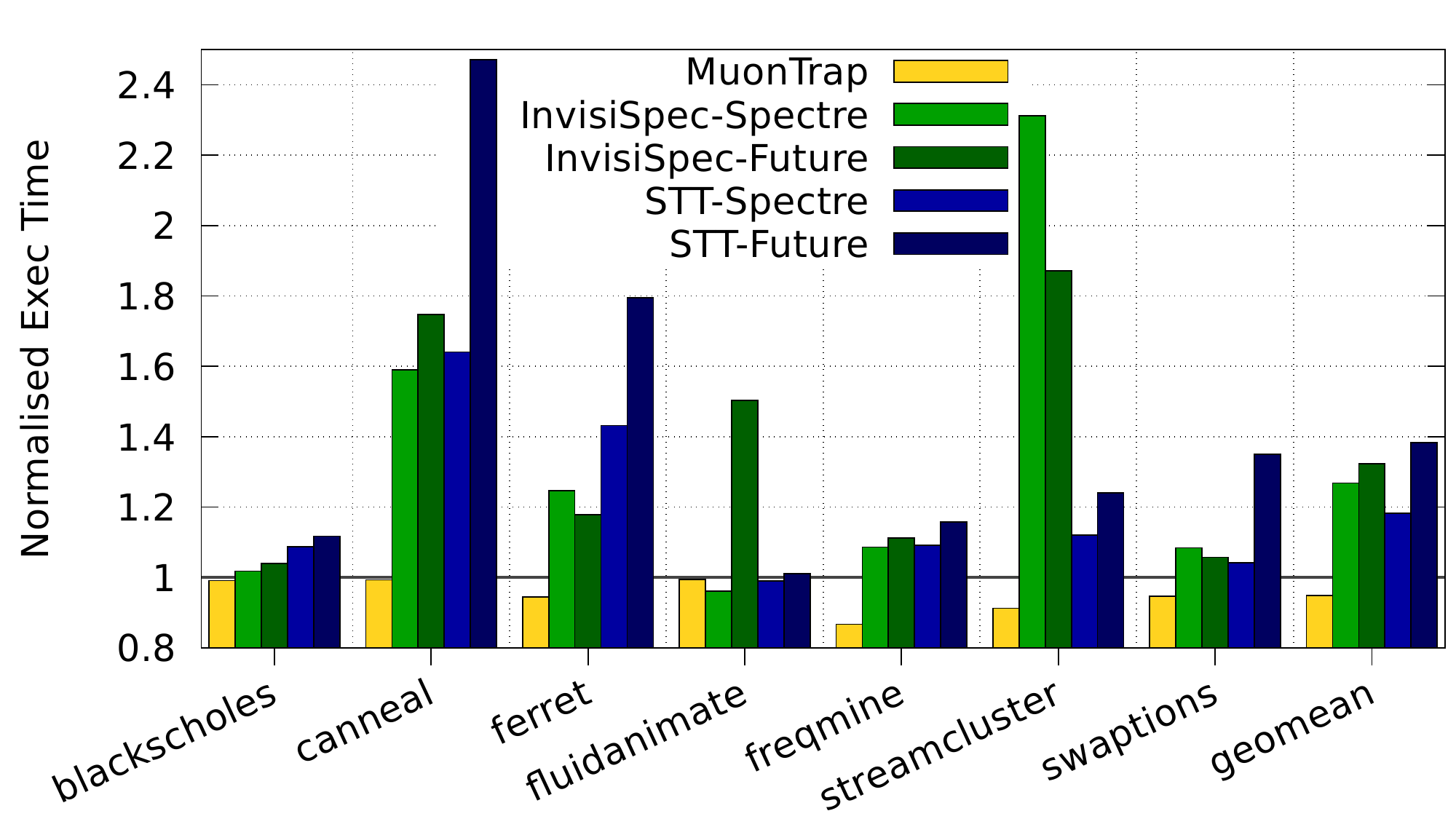}
		\vspace{-18pt}
	\caption{Normalised execution time from MuonTrap running Parsec, compared with InvisiSpec and STT.}
	\label{figs:parsec}
	\vspace{-15pt}
\end{figure}

\begin{table}
\small
\begin{tabulary}{\columnwidth}{lL}
\multicolumn{2}{c}{\textit{Main cores}}\\
\midrule
Core & 8-Wide, out-of-order, 2.0GHz \\
Pipeline & 192-Entry ROB, 64-entry IQ, 32-entry LQ, 32-entry SQ, 256 Int / 256 FP registers, 6 Int ALUs, 4 FP ALUs, 2 Mult/Div ALU \\
Tournament & 2048-Entry local,
8192-entry global,\\
Branch Pred.\ &  2048-entry chooser, 4096-entry BTB, 16-entry RAS \\
\vspace{-4pt}
\\
\multicolumn{2}{c}{\textit{Private core memory}}\\
\midrule
L1 ICache & 32KiB, 2-way, 1-cycle hit lat, 4 MSHRs \\
L1 DCache & 64KiB, 2-way, 2-cycle hit lat, 4 MSHRs \\
TLBs & 64-Entry, fully associative, split between instructions and data \\
Data \ucache & 2KiB, 4-way, 1-cycle hit lat, 4 MSHRs \\
Inst \ucache & 2KiB, 4-way, 1-cycle hit lat, 4 MSHRs \\
\vspace{-4pt}
\\
\multicolumn{2}{c}{\textit{Shared system state}}\\
\midrule
L2 Cache & 2MiB, 8-way, 20-cycle hit lat, 16 MSHRs, stride prefetcher \\
Memory & DDR3-1600 11-11-11-28 800MHz \\
OS & Ubuntu 14.04 LTS \\
Core count & 4 cores \\
\end{tabulary}
\vspace{-4pt}
\caption{Core and memory experimental setup.}
\label{tab:setuptable}
\vspace{-10pt}
\end{table}

\section{Evaluation}

\noindent We first look at overall performance for the whole MuonTrap technique on SPEC CPU2006 (4\% average slowdown) and Parsec (5\% average speedup), before comparing with related work re-evaluated on the same system~\cite{yan2018invisispec} and reported in the literature~\cite{STT-micro}. Then, we perform a tuning analysis on Parsec, focusing on cache size and associativity. We finally examine in detail the overheads from protections for the instruction cache, coherency protocol and prefetcher, \revision{A1}{and the optional clearing of the \ucache{} on every misspeculation}.
\label{sec:eval}

\subsection{Performance}

\noindent In \cref{figs:spec2006} we see that execution is slowed down by 4\% on average for SPEC CPU2006~\cite{SPEC2006} with MuonTrap.
Some workloads (povray, lbm) are sped up by virtue of the faster L0 data-cache access relative to an unprotected system.
Others are hampered by the small filter-cache size (bwaves), low associativity of the filter cache (cactusADM), the delayed commit-time prefetch mechanism (leslie3d and libquantum), the instruction filter cache (omnetpp) or a combination of all of these factors (zeusmp). The access delay for the L1 cache from being behind the filter cache also adds some overhead; geomean slowdown can be reduced further to 2\% if access is allowed in parallel, as we shall later see in \cref{fig:cacheextrasspec}. Still, for many workloads, the performance impact of all of these protections is negligible. Lbm is in fact sped up significantly using a filter cache; the in-order prefetching necessary for protection against speculative-execution attacks also allows the prefetcher to better pick up the access pattern in this workload, dramatically improves performance.

For Parsec on 4 cores (\cref{figs:parsec}), despite its protections a filter cache actually results in a speedup for each workload (geomean 5\%). This is because these parallel workloads are very amenable to having a small, 1-cycle L0 data cache in front of a conventional, slower, physically addressed 2-cycle L1, and the additional costs that clearing on context switches, on prefetches, on coherence, on evicting uncommitted data and on instruction fetch are not enough to outweigh this advantage. This is not the case for SPEC CPU2006, hence why systems that do not need the security of MuonTrap do not gain this performance benefit elsewhere.

\subsection{Versus InvisiSpec}
\label{ssec:invisispec}

\noindent InvisiSpec~\cite{yan2018invisispec} is a load-store-queue extension designed for the same purpose as the \ucache{} in this paper, in that it hides speculative execution.
However, we see in \cref{figs:spec2006} that MuonTrap typically has higher performance than either of InvisiSpec's two designs, despite the fact that MuonTrap also covers the instruction cache and prefetcher.
The first of these (InvisiSpec-Spectre) assumes that data can be made visible as soon as its speculative load is not dependent on any unresolved branches, introducing a slowdown of 9.7\%.
The second (InvisiSpec-Future) incurs a 18.5\% slowdown when it assumes that data is not safe to become visible until a load can no longer be squashed.
MuonTrap considers an instruction speculative until it commits, as this reduces hardware implementation complexity. This is close to the threat model of Invisispec-Future, and we often see performance spikes on similar workloads (bwaves, leslie3d, zeusmp); the cases where MuonTrap particularly suffers (cactusADM and leslie3d) are caused by the prefetcher, which is not protected by InvisiSpec.
However, MuonTrap still achieves lower performance impact than Invisispec-Spectre overall. 
On Parsec (\cref{figs:parsec}), the up to 2$\times$ performance disadvantage of both InvisiSpec-Spectre and InvisiSpec-Future is eliminated.

InvisiSpec stores cache lines in word-sized load-store log entries, whereas a MuonTrap speculative \ucache\ is a true cache, with cache-line-sized entries.
Our technique does not need to store multiple copies of data when there is temporal or spatial locality, reducing the amount of SRAM needed for the same coverage. 
InvisiSpec accesses do not participate in the cache coherence protocol at all, and so all loads must be replayed and potentially reload their data when the instruction commits, causing a delay at the end of the pipeline.
Our approach instead keeps \ucacheh\ data coherent, but prevents data leakage by delaying execution of instructions that would invalidate other caches with the same data, which is an uncommon case. 
A speculative \ucache\ must issue a second coherence request to gain exclusive access, much like the second request in InvisiSpec.
However, because a speculative \ucache\ is coherent, this does not delay commit, prevents load instructions from ever having to be re-executed, and does not require the cache line itself to be reloaded from the cache hierarchy.
This second cache-system access is asynchronous in a speculative \ucache, unlike in InvisiSpec, improving performance significantly. This coherency does mean that exclusive upgrades for stores must invalidate \ucache s. However, we see in  \cref{fig:specucacheinvalidatewrites} that this is typically a rare occurrence, as most stores are to data already in private caches. This trades off an expensive common-case operation (repeat memory accesses before a load can commit) for a rarer one (\ucacheh\ broadcasts for a subset of stores).


\subsection{Versus STT}

\noindent Speculative Taint Tracking~\cite{STT-micro}, similar to NDA~\cite{NDA-micro}, Conditional Speculation~\cite{CondSpec} and SpecShield~\cite{8714070}, is based on the policy of restricting the forwarding of speculative load values. For some compute-bound workloads in SPEC CPU2006 (\cref{figs:spec2006}), this adds little overhead. However, those with more complex memory-access patterns, such as astar and omnetpp, suffer high performance losses that MuonTrap can alleviate; on Parsec (\cref{figs:parsec}) we see geomean overheads of 18\% for STT-Spectre, and 38\% on the less permissive STT-Future, compared with 5\% speedup for MuonTrap. 

\subsection{Tuning Parameters}

\noindent We now look at how tuning the parameters of a \ucache\ within just the data hierarchy affects performance, using the Parsec benchmark suite running with 4 threads.

\myparagraph{Cache size}
\label{sec:eval:size}
\begin{figure}
\includegraphics[width=\columnwidth]{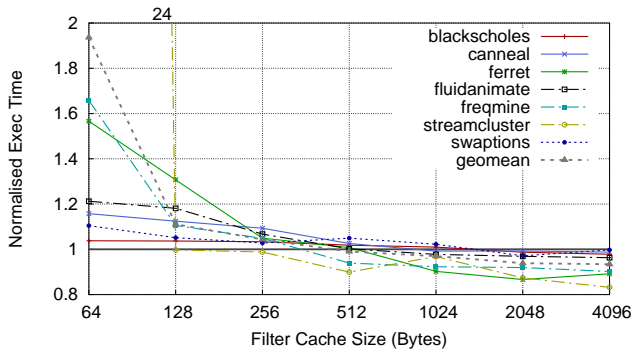}
\vspace{-15pt}
\caption{Performance of a fully associative \ucache\ added to the system in \cref{tab:setuptable}, with varying size, on Parsec.}
\label{fig:cachesize}
\vspace{-15pt}
\end{figure}
\noindent \Cref{fig:cachesize} shows normalised execution time for Parsec from adding a fully associative \ucache\ to the system at various sizes, normalised to the same system without any \ucache\ or associated protections. We see that, for some benchmarks, even a single cache line (64 bytes) is enough to get close to the performance of an unprotected system. This is because these workloads feature either high spatial and temporal locality, or little memory-level parallelism, and therefore early eviction of cache lines from the \ucache\ before they can be committed from execution is rare. For other workloads, however, particularly streamcluster and freqmine, enormous slowdowns are observed when the \ucache\ is too small.
However, these large slowdowns disappear for all benchmarks with four cache lines (256 bytes) of space in the cache or more. This is only a quarter of the number of loads from independent cache lines the processor we simulate can support (16 load elements), and yet it is sufficient to the point of producing a minor speedup. In cases where loads come from many different cache lines at once, there is little spatial or temporal locality available, and thus not writing back the data to the L1 does not affect performance.
We find that the majority of performance improvement can be attained with a 2,048 byte \ucache, giving no slowdown for any benchmark and a speedup of 6.9\% on average.
For all other experiments, we assume a \ucache\ of this fixed size, which comes at negligible area overhead compared with the core itself.

\myparagraph{Associativity}
\begin{figure}
\includegraphics[width=\columnwidth]{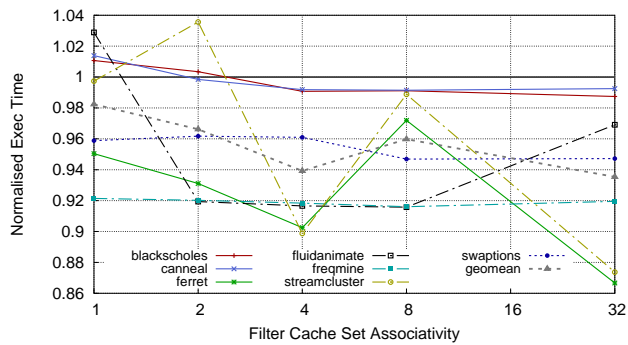}
\vspace{-15pt}
\caption{Normalised execution time when varying the associativity of a 2,048 byte (32 cache line) \ucache\ from direct mapped, to set associative, to fully associative, on Parsec.}
\label{fig:cacheassoc}
\vspace{-15pt}
\end{figure}
\noindent In \cref{fig:cacheassoc}, we reduce the full associativity (32-way) of the 2,048 byte \ucache\ in \cref{fig:cachesize} to show the performance impact. Some workloads (blackscholes, canneal, fluidanimate, and streamcluster) can be affected by conflict misses on \ucacheh\ data before it can be committed, and written back to the L1 cache. We set the associativity to 4-way to trade off access complexity compared to full associativity without reducing performance.

\begin{figure}[t]
\centering
\includegraphics[width=\columnwidth]{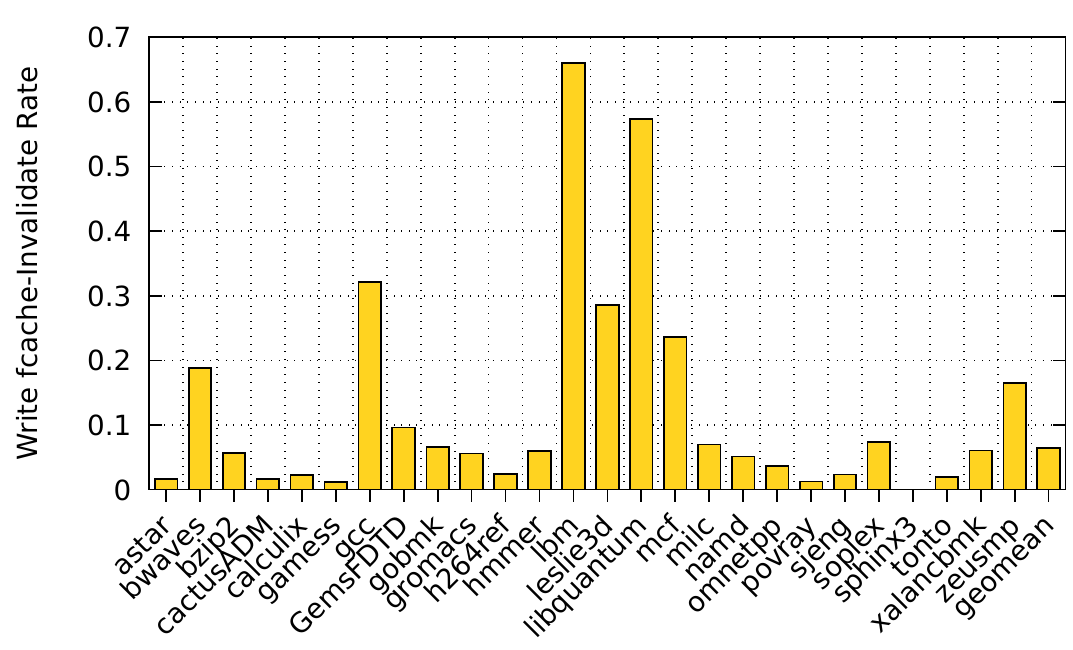}
\vspace{-15pt}
\caption{Proportion of writes that trigger \ucacheh\ invalidates for SPEC2006 under MuonTrap.}
\label{fig:specucacheinvalidatewrites}
\vspace{-15pt}
\end{figure}

\subsection{Cost Breakdown}

\noindent A \ucache\ on the core's data side on its own would be sufficient to protect against the original Spectre attack~\cite{Kocher2018spectre}. However, to protect against similar attacks on other parts of the memory system, MuonTrap adds in a variety of further mechanisms to cover coherency, instructions and prefetching. We now consider to what extent each of these contributes to MuonTrap's overheads. Each of these is already included in \cref{figs:spec2006} and \cref{figs:parsec}, but here we split them out for the most relevant workloads in each case to show the cost of each, looking at the most affected benchmark suite in each case. We finally show the further performance of a more intensive scheme that accesses the data \ucache\ and L1 in parallel, providing the potential to eliminate overheads further at the expense of complexity, which can reduce SPEC CPU2006 overhead from 4\% to 2\%.

\begin{figure}[t]
\includegraphics[width=\columnwidth]{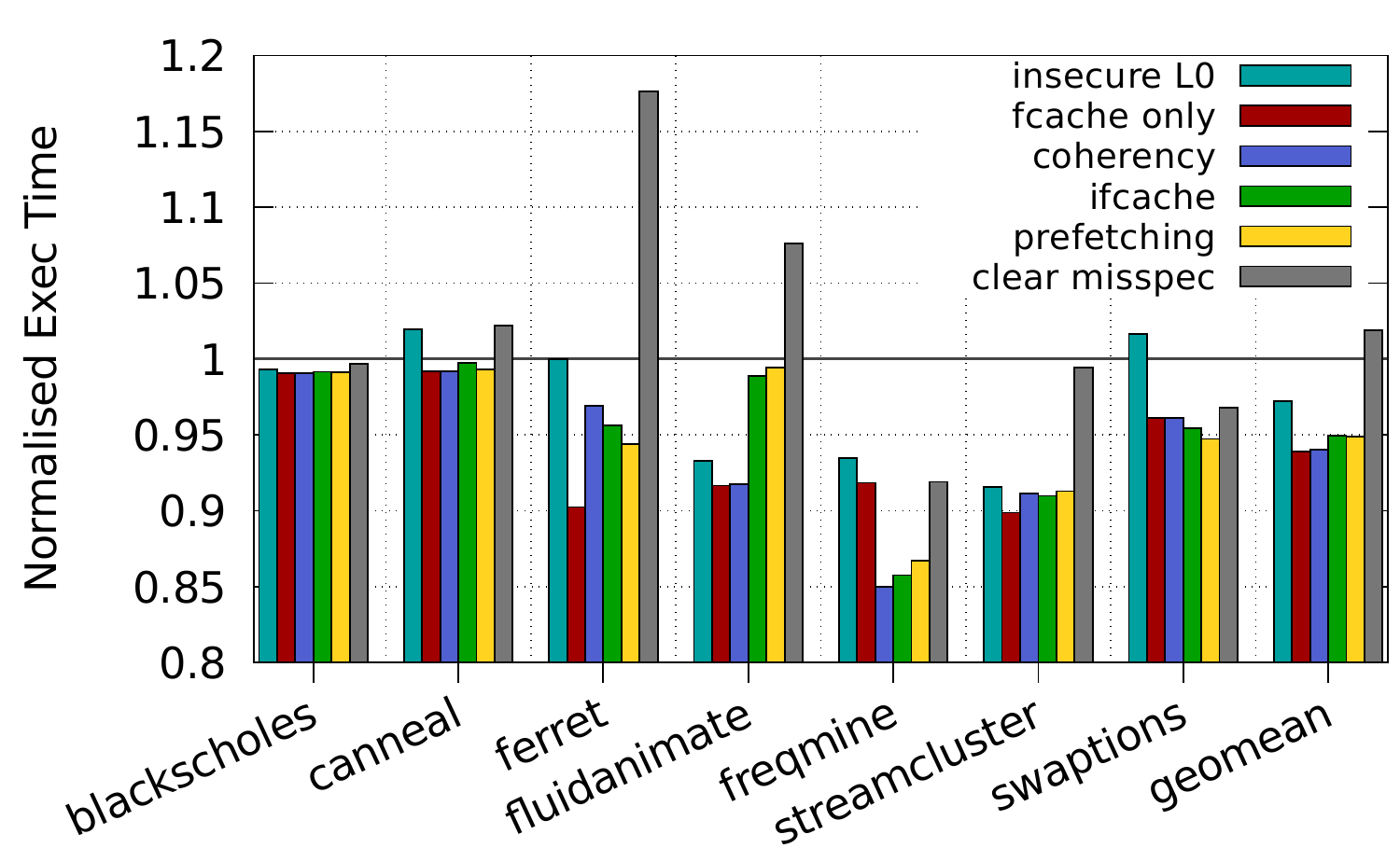}
\vspace{-18pt}
\caption{Normalised execution time from cumulatively adding protection mechanisms to the system on Parsec, compared with an unprotected system without \ucache s.}
\label{fig:cacheextras}
\vspace{-15pt}
\end{figure}

\begin{figure*}[t]
\includegraphics[width=2\columnwidth]{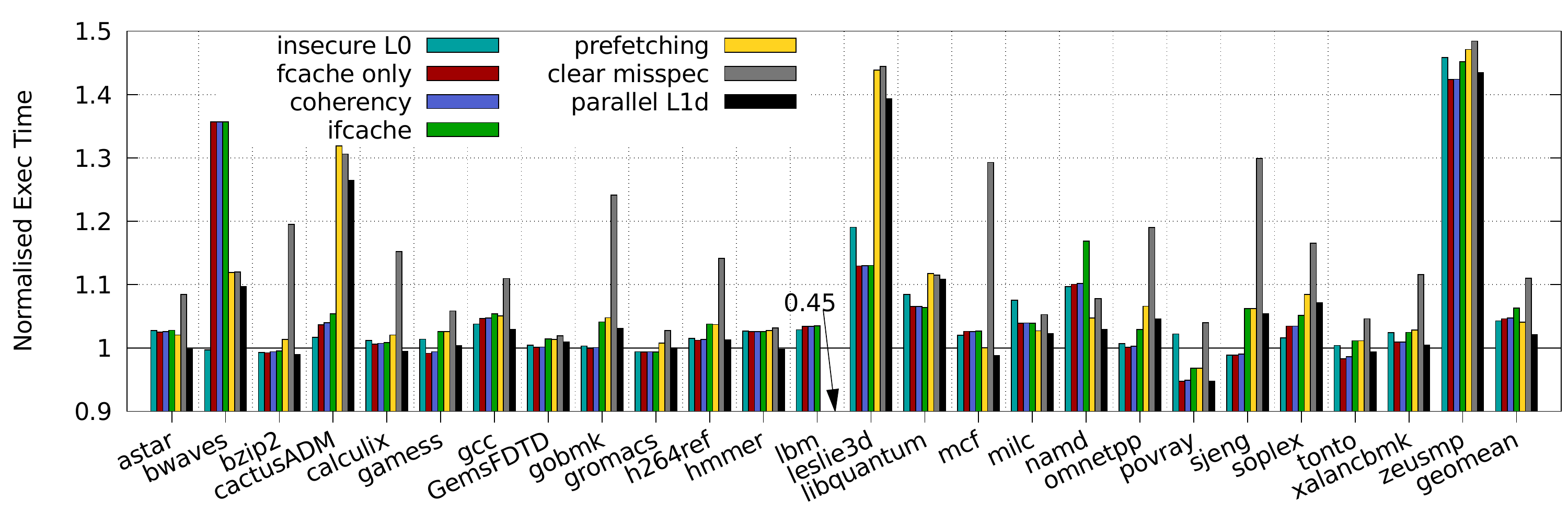}
\vspace{-12pt}
\caption{Performance from cumulatively adding protection mechanisms on SPEC CPU2006, relative to a filter-cache-free system.}
\label{fig:cacheextrasspec}
\vspace{-18pt}
\end{figure*}

Graphs for these additions are given in \cref{fig:cacheextras} for Parsec, and \cref{fig:cacheextrasspec} for SPEC CPU2006.
The following sections add mitigations successively, comparing the performance.

\myparagraph{Filter Cache Protection}
\revision{A3/D2}{An insecure L0 cache in place of the filter cache lowers performance on SPEC compared with the L1-only baseline, while improving it on Parsec. This is because workloads such as leslie3d and zeusmp react negatively to the increase in L1 latency, and have low hit rates in the L0. Meanwhile others, such as bwaves, only suffer once the filter cache protections are in place, where speculative state not propagating to the L1 causes performance loss.}

\myparagraph{Coherence Protection} The removal of speculative coherence state changes, described in \cref{sssec:coherenceprevention} (coherency in \cref{fig:cacheextras}) typically have negigible impact on the single-threaded SPEC workloads, and only a minor impact on the multi-threaded Parsec workloads. This is because speculative coherency state changes to other caches are relatively rare, and under normal operation a bus transaction takes long enough that delaying it to make it nonspeculative does not alter performance significantly. The upgrades to exclusive from shared when a load commits have minimal impact on contention and pipeline execution time, since coherence requests are already rare, and since they do not block commit as they can be done asynchronously. The two exceptions to this are ferret and streamcluster, and even there the speed of the \ucache\ outweighs the additional slowdown for limited coherency speculation.
Still, we have seen in \cref{fig:specucacheinvalidatewrites} that some workloads (bwaves, gcc, lbm, libquantum, mcr and zeusmp) trigger many write-miss broadcasts, and while we cannot uniquely isolate this overhead from other \ucache\ overheads, we see that several of these workloads do suffer some performance overhead as a result (\cref{fig:cacheextrasspec}).

\myparagraph{Instruction \UCache} Unlike with the data cache, adding an instruction \ucache{} (ifcache in \cref{fig:cacheextras}) rarely improves performance.
This is because the baseline instruction cache already has a 1-cycle access latency, and so adding a \ucache{} has no improvement potential, except for preventing some conflict misses, due to being more associative than the L1.
Indeed, slight performance drops occur with several benchmarks due to more lines being accessed before instruction commit than the instruction \ucache{} can store.
On Parsec (\cref{fig:cacheextras}) this is typically minor, save for fluidanimate, but even then the additional overhead is not enough to cause MuonTrap to lower performance overall. However, for SPEC CPU2006 (\cref{fig:cacheextrasspec}), we see multiple workloads taking a minor hit, and namd and sjeng taking a more significant penalty. 

\myparagraph{Commit-Time Prefetching}
Altering prefetching to be performed at commit time (\cref{ssec:prefetcher}, prefetching in \cref{fig:cacheextras}) instead of immediately after a load has been issued, is a double-edged sword. While Parsec is minimally affected, in \cref{fig:cacheextrasspec} for SPEC CPU2006 we see that while lbm and bwaves see significant performance improvement, through better tracking of the load stream without misspeculation, cactusADM and leslie3d suffer from reduced prefetch timeliness.

\myparagraph{Clear on Misspeculation}
\revision{A1}{Workloads that require within-process isolation of user-space sandboxes, and feature no further software mitigations for this use case, can configure MuonTrap to flush on all misspeculation. This increases overheads to 11\% for SPEC and 2\% on Parsec, as several workloads make use of both misspeculated data, and old accesses in the L0.}

\myparagraph{Parallel L1 and L0 access}
While Parsec has high performance regardless of \ucacheh\ configuration, the 4\% slowdown for SPEC CPU2006 leaves room for improvement. Even without additional protections, a \ucache\ alone is a significant detriment to many workloads simply because the \ucache\ cannot be sized large enough to cover any significant proportion of the working set, meaning that the additional delay to the L1 cache lowers performance. To mitigate this, at the expense of complexity, high-performance systems can be configured to access the 2-cycle L1 non-speculative cache and 1-cycle L0 \ucache\ simultaneously. We see in \cref{fig:cacheextrasspec} that this can reduce overhead from the 4\% reported in the rest of the paper, down to 2\%.

\subsection{Summary}

\noindent MuonTrap fact achieves a 4\% slowdown on SPEC CPU2006 (worst case 47\%), and a 5\% performance improvement on Parsec, compared to an insecure baseline. Since MuonTrap typically avoids limiting speculation within a threat domain, it generally sees overheads lower than existing techniques in the literature~\cite{yan2018invisispec,STT-micro}. While mechanisms such as in-order prefetching do have a performance impact, slowdowns are still typically small, and can be mitigated further by accessing \ucache s and non-speculative caches in parallel.

\section{Related Work}

\subsection{Speculative Side-Channel Attacks}

\noindent Spectre~\cite{Kocher2018spectre} allows the leakage of secret information from a victim process by attacking the branch predictor or poisoning the branch target buffer. Meltdown~\cite{Lipp2018meltdown} defeats kernel protections on some out-of-order superscalar processors due to a speculative side channel caused by not checking permissions on cache fill.  Other variants have also been proposed and implemented, including SpectrePrime~\cite{SpectrePrime}, which uses the cache-coherence system as a side channel for the same exploit. Spectre variants 1.1 and 1.2~\cite{kiriansky2018speculative} exploit the branch predictor as the attack mechanism but use speculative stores to shape that execution further, and variant 4~\cite{variant4} exploits speculative execution to leak data that should have been zeroed in program order. As these use cache side channels to leak data, our protection mechanism prevents their utilisation.


\subsection{Current Mitigations}

\myparagraph{Deployed Mitigations}
A variety of software mitigations exist for protecting against some variants of Meltdown and Spectre. For the former, kernel page-table isolation~\cite{kaiser} can be used to separate out kernel and user-space memory into separate address spaces, at overheads of up to 30\%~\cite{kaiser}. Spectre mitigations are also available in software, but as it is more difficult to protect against, mitigations tend to be more ad hoc: typically, program recompilation is required, and coverage is often low. Google's Retpoline~\cite{retpoline} replaces vulnerable branches with unconditional jumps, to override the branch-target buffer's predictor to prevent speculation. This defends against variant 2 of Spectre on x86, but is ineffective for Arm systems~\cite{armfaq}. For variant 1, on Arm systems non-speculative load barriers~\cite{armnonspec} can be inserted by the programmer to prevent particularly vulnerable loads from being exploited, however, this requires security knowledge by the programmer, reduces performance, and suffers from a lack of general coverage. LFENCE instructions can be used on x86 architectures~\cite{intelnfo} to similar effect, with similar downsides. Microcode updates such as Indirect Branch Restricted Speculation (IBRS)~\cite{intelnfo} also target variant 2 on Intel machines, by reducing speculation on kernel entry for indirect branches, but these cover only a subset of variant 2 exploits. 
New Intel~\cite{intel} and Arm~\cite{armv85aupdates} hardware designs feature variant 2 mitigations by isolating the branch-target buffer, preventing direct training by an attacker, but as variant 1 typically occurs by causing the victim to train itself, this strategy is of no benefit for the latter. 

\myparagraph{Memory Hierarchy}
InvisiSpec~\cite{yan2018invisispec}, as discussed in \cref{ssec:invisispec}, covers speculative execution by associating load-store-queue entries with cache lines, and repeating accesses to bring them into the cache when the instruction becomes nonspeculative. SafeSpec~\cite{DBLP:journals/corr/abs-1806-05179} also stores speculative data in fully-associative shadow structures attached to the load-store queue. It requires strict limits on the forwarding of data to be secure, which also means that it must be significantly overprovisioned, and that the same data must be able to exist multiple times in the same SafeSpec shadow structure to prevent side channels. The paper does not consider coherence-protocol attacks, and its structure likely prevents application of the coherence strategy applied to MuonTrap, since the invalidation operation is infeasibly expensive on the large, fully-associative structures necessary for SafeSpec to work. DAWG~\cite{kirianskydawg} uses dynamic partitioning to isolate cache entries in the absence of shared memory to prevent general cache side channels, but this is not feasible for cross-process Spectre attacks, where a large number different processes may be executing concurrently with mutual distrust.

CleanupSpec~\cite{Saileshwar:2019:CUA:3352460.3358314} allows speculative state to propagate through the memory system, using rollback techniques to undo changes to the caches. While this prevents the direct channel from reading the caches once this rollback is complete, the rollback mechanism is itself timing-dependent on secrets brought in by an attacker, unlike MuonTrap's constant-time invalidation of its \ucache s. Likewise, as CleanupSpec does not clear speculative state between protection domains, an attacker with code running concurrently with the victim's execution, but before it in program order, can observe state altered by the victim's execution.

\myparagraph{Load-Propagation Restriction}
SpecShield~\cite{8714070}, NDA~\cite{NDA-micro}, Conditional Speculation~\cite{CondSpec}, Sakalis et al.~\cite{SDVP} and Speculative Taint Tracking~\cite{STT-micro} are approaches that restrict various proportions of instructions dependent on speculative loads. While this can prevent the installation of secrets within a wide variety of side-channel mechanisms, and for workloads such as SPEC CPU2006 can be achieved with minimal slowdown~\cite{8714070,STT-micro}, workloads with more complex memory behaviour such as Parsec~\cite{STT-micro} and SPEC CPU2017~\cite{NDA-micro} suffer from the  limitations of load restriction regardless of how permissive. Taram et al.~\cite{taram2019context} present context-sensitive fencing, where memory fences are dynamically inserted into code streams by the microcode engine, to protect against kernel attacks.

\subsection{Side-Channel Attacks and Prevention}

\noindent Side channels exist more generally~\cite{SideChannelsPracticalLLC,Chen:2010:SLW:1849417.1849974,Gruss:2016:PSA:2976749.2978356} than the speculative attacks we focus on. In particular, common cryptographic-algorithm implementations~\cite{Kocher:1996:TAI:646761.706156,Osvik:2006:CAC:2117739.2117741}  can be vulnerable to leaking information about their input and secret keys if not designed to be timing independent on their input data. Related to side channels are covert-channel attacks~\cite{Percival05cachemissing,Chen:2014:ADC:2611765.2611766}, where two cooperating processes, one at higher security clearance than the other, modulate a shared resource to violate a mandatory access-control policy, to leak information. 
 
It is possible to prevent all side-channel attacks in hardware~\cite{Liu:2014:RFC:2742155.2742176,ZhenghongWang:2008:NCA:1521747.1521781,Domnitser:2012:NCL:2086696.2086714}. However, this involves modification of the entire cache hierarchy. To prevent speculative side-channel attacks, it is possible to modify only the level closest to the CPU, and still achieve strong security properties.
 



\subsection{\Ucache{}s}

\noindent Caches at the L0 level in a system have been utilised previously for power and performance~\cite{Kin:1997:FCE:266800.266818}. Duong et al.~\cite{Duong:2012:RLC:2380403.2380435} use small caches to reduce hit energy and conflict misses in embedded systems.  Tang et al.~\cite{1035013} use small instruction caches to reduce power consumption. In terms of industrial implementation, Qualcomm Krait processors ~\cite{AnandtechKrait} feature small,  single-cycle L0 caches. Similarly, many Arm processors use small microTLBs close to the core to achieve high performance even with physically-mapped L1 caches~\cite{microTLB}.

\section{Conclusion}

\noindent In this paper, we have shown that it is possible to mitigate speculative side-channel exploits between domains at low overheads in hardware, without removing speculation, by adding speculative \ucache s to store vulnerable state. In fact, MuonTrap can improve performance compared to an unmodified system: for the Parsec benchmark suite, performance is improved by 5\%, while SPEC CPU2006 is reduced by only 4\%. The modifications we have described to conventional systems to protect against Spectre~\cite{Kocher2018spectre} and its derivatives are simple to add to conventional systems, and provide strong performance and wide coverage.

\section*{Acknowledgements}
\noindent This work was supported by the Engineering and Physical Sciences Research Council (EPSRC), through grant references EP/K026399/1, EP/P020011/1 and EP/M506485/1, and ARM Ltd.
Additional data related to this publication is available in the repository at \url{https://doi.org/10.17863/CAM.50489}.



\setlength{\bibsep}{1.5pt plus 0.3ex}
\footnotesize
\bibliographystyle{abbrv}
\bibliography{refs}

\end{document}